\documentclass[9pt,twocolumn,twoside]{pnas-new}

\usepackage{lipsum}
\usepackage{bm}                                 
\usepackage{mathtools}                          
\usepackage{subcaption}
\usepackage{siunitx}

\usepackage{xr}
\makeatletter

\newcommand*{\addFileDependency}[1]{
\typeout{(#1)}
\@addtofilelist{#1}
\IfFileExists{#1}{}{\typeout{No file #1.}}
}\makeatother

\makeatletter
\long\def\XR@test#1#2#3#4\XR@{%
  \let\XR@next\@gobbletwo
  \ifx#1\newlabel
    \let\XR@next\@firstoftwo%
  \else\ifx#1\@input
     \let\XR@next\@secondoftwo
  \fi\fi
   \XR@next{\newlabel{\XR@prefix#2}{#3}}{\edef\XR@list{\XR@list#2\relax}}%
  \ifeof\@inputcheck\expandafter\XR@aux
  \else\expandafter\XR@read\fi}
\makeatother

\templatetype{pnasresearcharticle} 

\newcommand{\deriv}[2]{\frac{\partial #1}{\partial #2}}

\newcommand{\bzero}{\bm 0}

\newcommand{\bef}{\bm f}

\newcommand{\bq}{\bm q}

\newcommand{\bx}{\bm x}

\newcommand{\bC}{\bm C}

\newcommand{\mH}{\mathcal H}

\newcommand{\mL}{\mathcal L}

\newcommand{\mT}{\mathcal T}

\newcommand{\mV}{\mathcal V}

\title{Automated discovery of reprogrammable nonlinear dynamic metamaterials}

\author[a]{Giovanni Bordiga}
\author[b,a]{Eder Medina}
\author[c,a]{Sina Jafarzadeh}
\author[d,a]{Cyrill B\"osch}
\author[b]{Ryan P. Adams}
\author[e,a]{Vincent Tournat}
\author[a,1]{Katia~Bertoldi}

\affil[a]{
John A. Paulson School of Engineering and Applied Sciences, Harvard University, Cambridge, MA 02138, USA
}
\affil[b]{
Department of Computer Science, Princeton University, Princeton, NJ 08540, USA
}
\affil[c]{
Department of Energy Conversion and Storage, Technical University of Denmark, DK-2800 Lyngby, Denmark
}
\affil[d]{
Institute of Geophysics, ETH Zurich, 8092 Zurich, Switzerland
}
\affil[e]{
Laboratoire d'Acoustique de l'Universit\'e du Mans (LAUM), UMR 6613, Institut d'Acoustique - Graduate School (IA-GS), CNRS, Le Mans Universit\'e, France}

\leadauthor{Bordiga}

\authorcontributions{
Author contributions:
G.B., E.M., V.T., and K.B. designed research;
G.B. wrote simulation and optimization codes;
S.J. designed fabrication process;
G.B. and S.J. performed fabrication;
G.B., V.T., and S.J. performed experiments;
G.B., C.B., and E.M. performed simulations and optimizations;
G.B., E.M., S.J., C.B., R.P.A., V.T., and K.B. analyzed data and wrote paper.
}
\authordeclaration{The authors declare no competing interest.}
\correspondingauthor{\textsuperscript{1}To whom correspondence should be addressed. E-mail: \href{mailto:bertoldi@seas.harvard.edu}{\texttt{bertoldi@seas.harvard.edu}}}


\begin{abstract}
Harnessing the rich nonlinear dynamics of highly-deformable materials has the potential to unlock the next generation of functional smart materials and devices.
However, unlocking such potential requires effective strategies to spatially design optimal material architectures for desired nonlinear dynamic responses such as guiding of nonlinear elastic waves, energy focusing, and cloaking.
Here, we introduce an inverse-design  framework for the discovery of flexible mechanical metamaterials with a target nonlinear dynamic response.
The desired dynamic task is encoded via optimal tuning of the full-scale metamaterial geometry through an inverse-design approach powered by a custom-developed fully-differentiable simulation environment.
By deploying such strategy, we design mechanical metamaterials  tailored for energy focusing, energy splitting, dynamic protection, and nonlinear motion conversion. 
Furthermore, we illustrate that our design framework can be expanded to automatically discover reprogrammable architectures capable of  switching between different dynamic tasks.
For instance, we encode two strongly competing tasks---energy focusing and dynamic protection---within a single architecture, utilizing static pre-compression to switch between these behaviors.
The discovered designs are physically realized and experimentally tested, demonstrating the robustness of the engineered tasks.
All together, our approach opens an untapped avenue towards designer materials with tailored robotic-like reprogrammable functionalities.

\end{abstract}

\dates{This manuscript was compiled on \today}

\begin{document}

\maketitle
\thispagestyle{firststyle}
\ifthenelse{\boolean{shortarticle}}{\ifthenelse{\boolean{singlecolumn}}{\abscontentformatted}{\abscontent}}{}

Precise control over the nonlinear behavior of material structures is essential for various dynamic tasks, including energy harvesting from elastic pulses, impact mitigation, and mechanical signal processing.
Mechanical metamaterials---artificially engineered materials with mechanical responses determined by structure rather than composition---have emerged as a promising platform to achieve such control.
By carefully arranging specially designed building blocks in space, these metamaterials have demonstrated complex functionalities such as focusing~\cite{ma_2016}, executing mathematical operations~\cite{zangeneh-nejad_2021,silva_2014}, and cloaking objects ~\cite{zhang_2011,stenger_2012,xue_2021}. 
However, most of the proposed designs operate in the linear regime and are optimized for a single functionality.

Recent advancements have highlighted the enormous potential of flexible metamaterials in controlling nonlinear waves \cite{deng_2021a,patil_2022}.
By leveraging their ability to undergo large deformations, exploit instabilities, and navigate multi-welled energy landscapes, these systems have demonstrated capabilities such as unidirectional signal propagation \cite{nadkarni_2016a}, long-range propagation even in the presence of dissipation \cite{raney_2016}, and impact mitigation \cite{yasuda_2019}.
Nevertheless, the exploration and rational control of nonlinear dynamics in flexible mechanical materials is still in its early stages, with most studies focused on periodic systems.
Designing periodic material structures inherently restricts our ability to manipulate energy flow through space and time.
While it has been shown that defects can lead to diverse transmission pathways in the underlying lattice~\cite{jin_2020}, the placement of these defects has relied on intuition rather than systematic approaches.
By transitioning to automated design strategies, it may be possible to unlock new performant two-dimensional architectures that can achieve precise control over energy flow in the nonlinear regime~\cite{zaiser_2023}.

Optimization is an attractive principle for automated design, and it has been successful in the identification of metamaterials with desired dynamic responses within the linear regime.
Specifically, gradient-based topology optimization has yielded metamaterials exhibiting maximized relative size of bandgaps~\cite{sigmund_2003a,liu_2020a}, negative effective properties~\cite{dong_2017,li_2017}, directional propagation~\cite{he_2018a}, topologically protected wave modes~\cite{bosch_2020}, minimal dynamic compliance~\cite{jensen_2007}, and optimal reflection and dissipation of ground-borne elastic wave pulses~\cite{jensen_2007a}.
Further, the gradient-based topology optimization approach has been extended to phononic crystals which are tunable through elastic deformation~\cite{evgrafov_2008,sharma_2022,dalklint_2022}.
However, there has been significantly less progress in the control of nonlinear waves~\cite{fraternali_2009}, with efforts focused on the use of genetic algorithms.
While gradient-free methods can easily explore the complex energy landscape typical of flexible metamaterials, they are not suitable for problems involving a large number of design parameters.
At the same time, the combination of large parameter spaces and gradient-based optimization has been shown to be a tremendous success in discovering underlying patterns and behaviors~\cite{silver_2018,raissi_2019,fawzi_2022}.
Within mechanics, gradient-based optimization has been demonstrated to be effective for the design of elasto-plastic structures for failure resistance under impact~\cite{akerson_2023} and for shape-morphing cellular solids~\cite{oktay_2023}.
Although these works indicate the great potential of gradient-based optimization for nonlinear problems, the rich space of dynamical behaviors in nonlinear mechanical metamaterials remains relatively unexplored.

\begin{figure*}[htb]
    \centering
    \includegraphics[width=\textwidth]{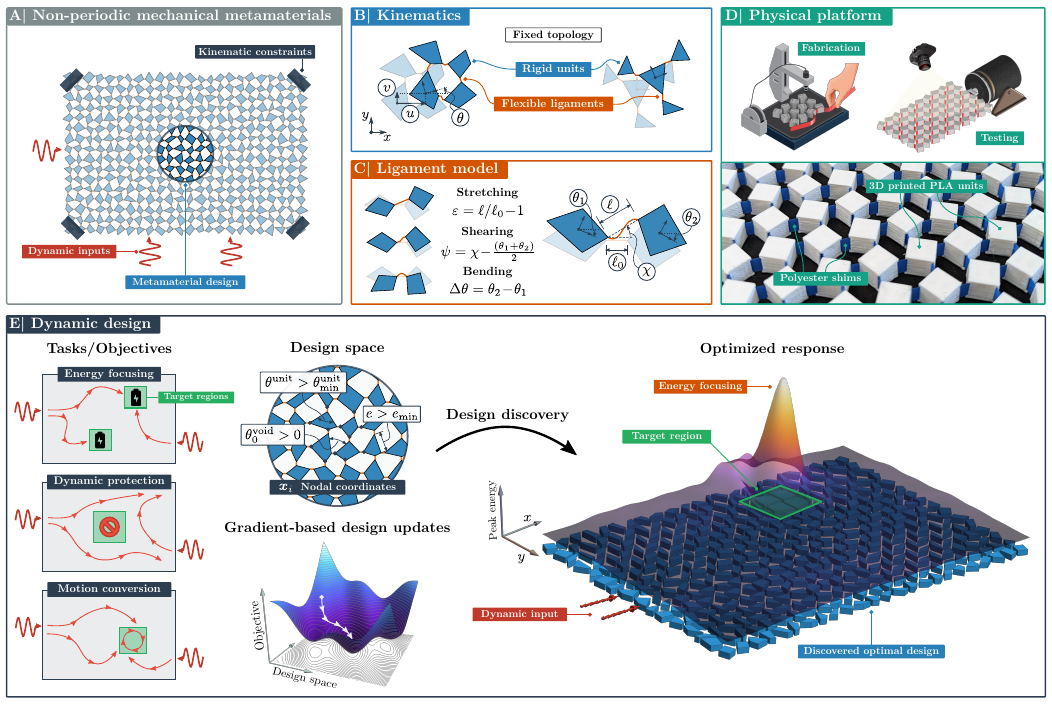}
    \caption{
        \textbf{Automated design of reprogrammable nonlinear dynamic metamaterials.}
        (\textbf{A}) We focus on 2D flexible mechanical metamaterials comprising a network of rigid units connected by flexible ligaments.
        (\textbf{B})-(\textbf{C}) Assuming a fixed topology for the connections, the dynamic behavior of the system is governed by (\textbf{B}) the rigid-body kinematics of the units and (\textbf{C}) the mechanical response of the flexible ligaments.
        (\textbf{D}) The metamaterials are physically realized via 3D-printing of PLA units connected by thin flexible plastic shims.
        (\textbf{E}) Given a high-level description of the dynamic tasks (e.g., energy focusing at target locations, protection, motion conversion), the design space of non-periodic geometries is explored efficiently through the use of adjoint gradients and a gradient-based optimizer.
    }
    \label{fig:design_framework}
\end{figure*}

Here, we introduce a framework to automate the design of flexible metamaterial structures that can execute desired nonlinear dynamic tasks and whose functionality can be reprogrammed on-the-fly.
In particular, we focus on 2D flexible mechanical metamaterials comprising a network of rigid units connected by flexible ligaments (see Fig.~\ref{fig:design_framework}A--D), which have gained significant attention due to their ability to exhibit effective negative Poisson's ratio~\cite{grima_2000,cho_2014}, a wide range of target static nonlinear mechanical responses~\cite{celli_2018,coulais_2018,czajkowski_2022,zheng_2023}, and support the propagation of solitary pulses~\cite{deng_2019b,yasuda_2020}.
By leveraging recent advancements in automatic differentiation~\cite{bradbury_2018, schoenholz_2020,minkov_2020,goodrich_2021}, we demonstrate how the nonlinear dynamic response of these metamaterials can be precisely tailored to execute complex tasks such as energy focusing, energy splitting, dynamic protection, and nonlinear motion conversion (Fig.~\ref{fig:design_framework}E).
Additionally, we extend the design framework to create architectures capable of seamlessly switching between different tasks.
As an example, we encode two strongly competing tasks---energy focusing and dynamic protection---within a single architecture and harness static pre-compression to switch between these behaviors.
To demonstrate the robustness of the engineered tasks, the discovered designs are physically realized and tested.
All together, the presented results highlight the efficacy of our framework in enabling the non-electronic encoding of reprogrammable nonlinear dynamic tasks in artificial material structures.

\section*{Design strategy}


To rationally design 2D flexible mechanical metamaterials with target nonlinear dynamic responses, we  developed a fully-differentiable simulation environment that leverages automatic differentiation (AD) tools~\cite{bradbury_2018,schoenholz_2020}.
This approach automates the derivation of the equations of motion from an energy functional and, therefore, greatly simplifies the modeling and simulation process.
Furthermore, it facilitates the inverse design of metamaterials by automatically computing gradients of the solution with respect to any design parameter.
Our design strategy consists of the following four steps.


\subsection*{Derivation and solution of equations of motion}
The response of 2D flexible mechanical metamaterials comprising a network of rigid units connected by flexible ligaments can be captured using a discrete model comprising rigid units connected at their vertices by a combination of elastic springs~\cite{deng_2017,coulais_2018}  (Fig.~\ref{fig:design_framework}A--C).
In particular, the strain energy of the $i$-th ligament is assumed of the form
\begin{equation}
\label{eq:ligament_energy}
    \mV_i = \frac{1}{2} \left[k_\ell (\varepsilon_i \ell^0_i)^2 + k_\theta {\Delta\theta_i}^2 + k_s (\psi_i\ell^0_i)^2\right]+\mV^\text{c}_i \,,
\end{equation}
where $k_\ell$, $k_\theta$ and $k_s$ denote the stiffness of the ligament upon stretching, bending, and shearing, respectively.
Moreover,~${\varepsilon_i=\ell_i/\ell^0_i-1}$, with $\ell_i$ and $\ell_i^0$ being the deformed and rest length of the $i$-th ligament, respectively,~${\Delta\theta_i=\theta^{(2)}_i-\theta^{(1)}_i}$, with $\theta^{(1)}_i$ and $\theta^{(2)}_i$ being the rotation of the two  units connected to the ligament and~${\psi_i=\chi_i-(\theta^{(1)}_i+\theta^{(2)}_i)/2}$, with~$\chi_i$ being the angle between the deformed and the undeformed configuration of the ligament (Fig.~\ref{fig:design_framework}C).
Finally, $\mV_i^\text{c}$ is a differentiable contact model in the form of a strain energy term accounting for the contact between the rigid units connected by the $i$-th ligament.
Such contact energy is assumed to be of the form
\begin{equation}
    \label{eq:contact_energy}
    \mV_i^\text{c} = \frac{1}{2} k_\text{c} \left(\theta_i^\text{void}-\theta_\text{cutoff}\right)^2 \left(1-\tau_i^2\right)^{-1} \mH(\theta_\text{cutoff}-\theta_i^\text{void})
    \,,
\end{equation}
where~${\theta_i^\text{void}=\Delta\theta_i+\theta^\text{void}_{i,0}}$ with $\theta^\text{void}_{i,0}$ being the rest void angle (Fig.~\ref{fig:design_framework}E), and $\mH$ denoting the Heaviside function.
Moreover, $k_\text{c}$ controls the initial stiffness of the contact while the factor $\left(1-\tau^2\right)^{-1}$, with~${\tau=(\theta^\text{void}-\theta_\text{cutoff})/(\theta_\text{cutoff}-\theta_\text{min})}$, introduces a vertical asymptote at~${\theta^\text{void}=\theta_\text{min}}$ to avoid compenetration.

The Lagrangian of a metamaterial comprising $N_\text{h}$ ligaments and $N_\text{u}$ rigid units can then be written as
\begin{equation}
    \mL = \mT - \mV = \sum_{i=1}^{N_\text{u}}\mT_i-\sum_{i=1}^{N_\text{h}}\mV_i 
\end{equation}
where $\mV_i$ is given by~\eqref{eq:ligament_energy}  and $\mT_i$ denotes the kinetic energy of the $i$-th rigid unit,
\begin{equation}
\label{eq:kinetic_energy}
    \mT_i = \frac{1}{2} \left( \rho A_i \left(\dot{u}_i^2 + \dot{v}_i^2\right) + \rho I_i \dot{\theta}_i^2 \right) \,,
\end{equation}
where $\dot{u}_i$, $\dot{v}_i$, and $\dot{\theta}_i$ are the horizontal, vertical, and rotational velocities of the center of mass of the $i$-th unit, and $\rho A_i$ and $\rho I_i$ denote the corresponding inertia (Fig.~\ref{fig:design_framework}B).
We then take advantage of AD to take the partial derivatives of $\mL$ with respect to all degrees of freedom of the metamaterial and obtain the equations of motion as
\begin{equation}
\label{eq:equations_of_motion}
    \frac{\text{d}}{\text{d}t} \deriv{\mL}{\dot{\bq}} - \deriv{\mL}{\bq} = \bef_\text{ext} \,.
\end{equation}
where ${\bq=\left\{u_1,v_1,\theta_1,\dots, u_{N_\text{u}},v_{N_\text{u}},\theta_{N_\text{u}}\right\}}$ is a vector collecting the displacement components of all $N_\text{u}$ units and $\bef_\text{ext}$ is the external force vector.
Such external loading function allows for modeling applied excitation as well as dissipation so that ${\bef_\text{ext}=\bef_\text{app}+\bef_\text{damp}}$.
For the results shown in this work, we assume displacement-driven loading conditions, hence $\bef_\text{app}=\bzero$.
Moreover, a simple linear viscous damping model is assumed so that $\bef_\text{damp}=-\bC\dot\bq$ with $\bC$ being a diagonal matrix with translation and rotation damping coefficients $c_u$ and $c_\theta$, respectively.
As~\eqref{eq:equations_of_motion} is a highly nonlinear system of ODEs, we numerically solve for the response $\bq(t)$ using a Dormand--Prince explicit solver with adaptive stepsize~\cite{dormand_1980} (see \href{SI-link}{Supporting Information}, Section~2A for more details).

%

We highlight that our simulation environment fully automates the derivation of~\eqref{eq:equations_of_motion} once all the energy functions~(\ref{eq:ligament_energy})--(\ref{eq:kinetic_energy}) and external forcing $\bef_\text{ext}$ are defined as a function of all the relevant inputs.
This approach provides the flexibility to change geometry, topology, and energy functions in a modular fashion.

\subsection*{Definition of the design space}
Given that the behavior of a mechanical metamaterial can be significantly altered by the shape of its rigid units~\cite{deng_2022a}, we choose here to use the shape of these rigid units as design space.
Such a design space can be parametrized via a vector $\bx$ collecting the coordinates of vertices of the $N_\text{u}$ rigid units in the undeformed (reference) configuration.
It is important to note that in order to guarantee identical mechanical response across all ligaments, the design space is constrained to maintain a constant rest length $\ell_0$.
The reference orientation of the ligaments is also kept constant during optimization.
These constraints are enforced for the experimental convenience of characterizing a single set of hinge parameters, and are not fundamental limitations of our design framework.
Furthermore, to both avoid infeasible designs and ensure manufacturability, we restrict the minimum edge length and vertex angle of the units, and constrain the void angle between neighboring units to be positive (see \href{SI-link}{Supporting Information}, Section~1 for additional details on the geometric parametrization and constraints).

\subsection*{Optimization}
Finally, we pose the question: How can we discover a geometry that can perform a desired dynamic task such as maximizing or minimizing the kinetic  energy at target locations?
By computing gradients with respect to the geometry parameter~$\bx$, it becomes possible to navigate a design space of much higher dimensionality than would be possible with a gradient-free method.
For each dynamic task, we specify an objective function $J(\bq(t),\bx)$ that reflects the efficacy of the geometry~$\bx$ in solving that task.
Given any design $\bx$, we numerically integrate~\eqref{eq:equations_of_motion} and then evaluate such objective function $J(\bq(t),\bx)$.
Further, we take advantage of AD to compute its gradient as
\begin{equation}
\label{eq:gradient_J}
    \frac{\text{d}J}{\text{d}\bx} = \deriv{J}{\bq}\deriv{\bq}{\bx} + \deriv{J}{\bx} \,,
\end{equation}
where the response $\bq(t)$ implicitly depends on the design $\bx$ by means of the constraint imposed by~\eqref{eq:equations_of_motion}.
Note that the gradient term $\partial J / \partial \bx$ is directly evaluated via AD, while the term~$\deriv{J}{\bq}\deriv{\bq}{\bx}$ is computed by solving the adjoint problem associated with~\eqref{eq:equations_of_motion} using the same time integrator~\cite{dormand_1980}.
Finally, $\text{d}J/\text{d}\bx$ is  passed to the optimizer to update the design $\bx$ until convergence.
In order to handle nonlinear constraints on the design space, required to ensure manufacturability (i.e., lower bounds on the edge lengths and angles of the units), we adopt the Method of Moving Asymptotes (MMA)~\cite{svanberg_1987} as the optimizer (provided by the NLopt library~\cite{johnson_2007}; see \href{SI-link}{Supporting Information}, Section~2A for more details).
This optimization algorithm can be readily applied to multi-task problems by adopting the simple approach of optimizing a convex linear combination (scalarization) of multiple objectives.
In particular, we can deploy this strategy to design reprogrammable architectures that can switch between multiple tasks.
In this way, we can discover a range of solutions that explore the trade-off between distinct functionalities, thus identifying the Pareto front~\cite{osanov_2016} (\href{SI-link}{Supporting Information}, Section~2B for more details).
As a result, our design framework allows us to efficiently explore the large space of non-periodic architectures and converge to a performant---even if only locally optimal---design that encodes one or multiple tasks.

\subsection*{Fabrication, characterization, and testing}
To evaluate the performance of the optimized designs, we fabricate them using 3D-printed polylactic acid (PLA) units and thin polyester plastic shims with rest length of~${\ell_0=\SI{2.3}{mm}}$ (Fig.~\ref{fig:design_framework}D).
The mass density used in the 2D model is determined by measuring the mass of a representative sample and dividing it by the area of the PLA units, obtaining ${\rho=\SI{6.18}{kg/m^2}}$.
The mechanical behavior of the hinges is systematically investigated by subjecting samples (comprising an array of $4\times4$ squares connected by these hinges) to tension, compression, and shear.
The experimental responses closely match the model predictions for ${k_\ell=\SI{120}{N/mm}}$, ${k_s=\SI{1.19}{N/mm}}$, and ${k_\theta=\SI{1.50}{N.mm}}$.
Furthermore, we find that the free oscillations of a square unit connected to the ground by a hinge align well with the model predictions for damping coefficients ${c_u=\SI{2.9e-2}{kg/s}}$ and ${c_\theta=\SI{1.2e-7}{kg.m^2/s}}$.
In addition, the contact stiffness is assumed to be ${k_\text{c}=k_\theta}$, and contact angles are chosen as ${\theta_\text{min}=\SI{-15}{\degree}}$ and ${\theta_\text{cutoff}=\SI{-10}{\degree}}$ to prevent excessive overlap (contact) between the units.
Finally, to ensure manufacturability, the minimum edge length and vertex angle of the units are set as ${e_\text{min}=\SI{3}{mm}}$ and ${\theta^\text{unit}_\text{min}=\SI{30}{\degree}}$, respectively.

The fabricated structures are dynamically excited using a low-frequency shaker and the response is recorded with a high-speed camera.
A digital image correlation (DIC)-based tracking algorithm is then used to reconstruct the displacement field of all the units and compare it to the simulated response (see \href{SI-link}{Supporting Information}, Section~4 and~5 for more details on fabrication and experimental methods).

\section*{Results}
\subsection*{Energy focusing}
\begin{figure*}[htb]
    \centering
    \includegraphics[width=\textwidth]{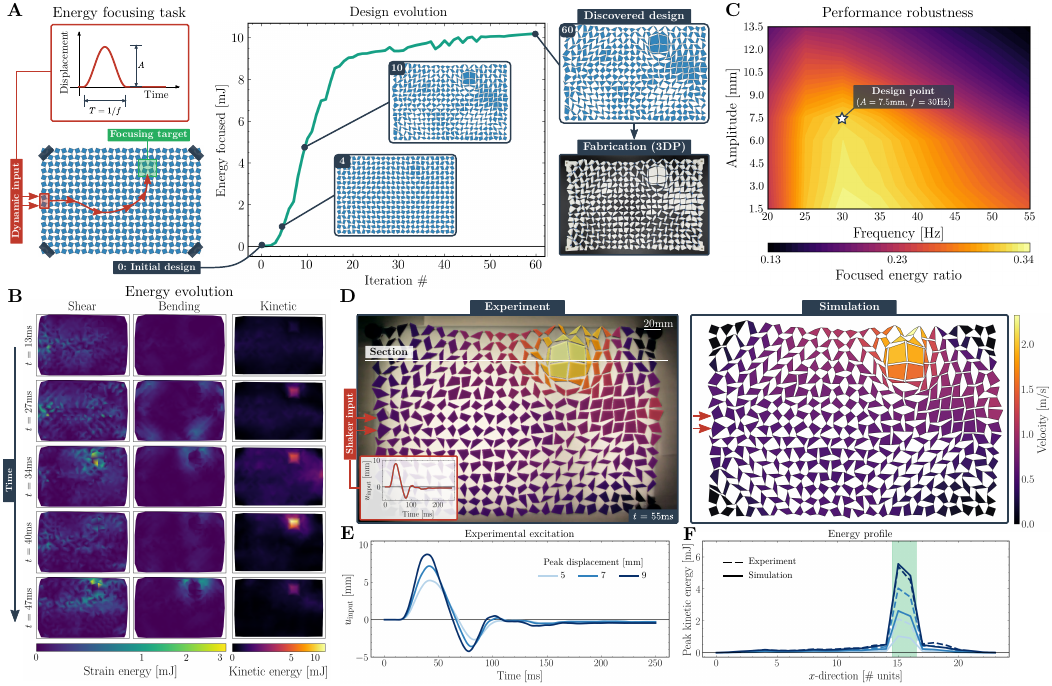}
    \caption{
        \textbf{Energy focusing.}
        (\textbf{A}) The dynamic task of focusing the energy provided by a dynamic excitation on the left boundary of the system towards a target location is encoded in the metamaterial structure through our gradient-based design strategy.
        The design evolution during optimization shows the effectiveness of the method in identifying the best direction in the huge design space of possible architectures (results shown for an input pulse $A=0.5\,s=\SI{7.5}{mm}$, $f=\SI{30}{Hz}$).
        (\textbf{B}) Simulation snapshots of the spatial distribution of the most relevant energy components (shear, bending, and kinetic) at $t=13,\,27,\,34,\,40$ and \SI{47}{ms}.
        (\textbf{C}) Contour plot of the simulated focused energy ratio (integrated kinetic energy of the target region compared to elsewhere, i.e. $J_{\Omega_\text{t}}/J_{\Omega\setminus\Omega_\text{t}}$) for the optimized design as a function of input frequency and amplitude predicting robustness with respect to the excitation away from the design point.
        (\textbf{D}) Experimental and numerical snapshot of the velocity field at  $t=\SI{55}{ms}$ for the optimized structure subject to the input shown in the inset.
        (\textbf{E}) Additional experimental excitation signals used in the experiments and (\textbf{F}) corresponding evolution of the peak kinetic energy along the white line in panel D.
    }\label{fig:energy_focusing_single_input_single_target}
\end{figure*}

To demonstrate our optimization-based approach, we first seek a metamaterial design that directs the kinetic energy provided by a large amplitude pulse towards a target region $\Omega_\text{t}$ (Fig.~\ref{fig:energy_focusing_single_input_single_target}A).
To achieve this goal, we maximize the time-integral of the kinetic energy at $\Omega_\text{t}$
\begin{equation}
    \label{eq:objective_focusing}
    J_{\Omega_\text{t}}(\bx) = \sum_{i\in\Omega_\text{t}} \int_0^{t_\text{f}} \mT_i(\bq(t),\bx)\,\text{d}t \,,
\end{equation}
upon application of a pulse-like excitation on the left edge of the domain (see region highlighted in red in Fig.~\ref{fig:energy_focusing_single_input_single_target}A).
In particular, we focus on a domain comprising $24\times16$ units, choose $\Omega_\text{t}$ to be four blocks located in the upper right part of the domain (see region highlighted in green in Fig.~\ref{fig:energy_focusing_single_input_single_target}A) and consider the following excitation signal
\begin{equation}
    \label{eq:pulse_excitation}
    u_\text{input}(t) = \frac{A}{2} \left(1-\cos(2\pi f t)\right) \mH(1/f-t)\mH(t) \,,
\end{equation}
where $A=\SI{7.5}{mm}$ and $f=\SI{30}{Hz}$ control the amplitude and width of the single pulse applied to the structure.
Note that we choose the upper limit of integration in~\eqref{eq:objective_focusing} to be $t_\text{f} = 2/f$, since we have found that larger values of integration time lead to very comparable performance, while requiring a higher computational cost (\href{SI-link}{Supporting Information}, Fig.~S8).

As shown in Fig.~\ref{fig:energy_focusing_single_input_single_target}A, our optimization algorithm quickly alters the initial design comprising squares with a center-to-center spacing $s=\SI{15}{mm}$ and bias angle $\theta_0=\SI{25}{\degree}$.
Specifically, the algorithm modifies the geometry throughout the entire domain, and after approximately 60 iterations identifies a design that concentrates $J_{\Omega_\text{t}}/t_\text{f}\approx\SI{10}{mJ}$ of energy in the target area.
This represents a more than 400-fold enhancement compared to the initial periodic design, which focused around $\approx\SI{0.02}{mJ}$ in the target area.
The optimized design features large quadrilaterals within the target region, surrounded by smaller units with a high aspect ratio, creating a structure that resembles an elastic resonator made of a high-inertia region embedded in a softer surrounding environment~\cite{liu_2000}.
Additionally, a high-density region is situated just below the target and low-density regions are positioned near the four corners of the domain.
To understand the roles played by these regions, we analyze the temporal evolution of bending, shear, and kinetic energies over this optimized design (Fig.~\ref{fig:energy_focusing_single_input_single_target}B).
At $t=\SI{13}{ms}$, we find that the applied input primarily causes shearing of the ligaments in a straight region ahead of the excitation point.
By $t=\SI{34}{ms}$, the applied energy has transferred into both shear and bending energies and the kinetic energy begins to focus at the target location.
During this energy exchange, the high-density region below the target acts as a high-inertia reflection area where the kinetic energy gets distributed before being transferred towards the target region (Movie~S1).
This process continues until most of the kinetic energy is concentrated at the target, and very low strain energy is present elsewhere, resulting in a significant focusing event at the desired location ($t=\SI{40}{ms}$).
The exchange mechanism between the kinetic energy in the target area and the strain energy in the rest of the domain continues until all the energy is dissipated.


Next, we investigate the robustness of the optimized design with respect to different input pulses.
Towards this end, we simulate its response upon application of inputs with $A\in[0.1,0.9]s=[1.5, 13.5]\,\si{mm}$, $f\in[20, 55]\,\si{Hz}$.
In Fig.~\ref{fig:energy_focusing_single_input_single_target}C we report the integral of the energy focused on the target region compared to the rest of the domain, $J_{\Omega_\text{t}}/J_{\Omega\setminus\Omega_\text{t}}$, for all the considered inputs.
Notably, we observe that comparable focusing performance is attained for inputs within a substantial region surrounding the input  considered in the optimization (see also \href{SI-link}{Supporting Information}, Section~3A for more analysis on robustness).
This robustness with respect to different excitation signals is a key prerequisite for experimental validation.

To experimentally demonstrate the focusing task, we fabricate the optimized design and dynamically excite it using a low-frequency shaker.
In these experiments, the input excitation is extracted via DIC tracking of the units connected to the shaker and then fed into our simulations for comparison.
In Fig.~\ref{fig:energy_focusing_single_input_single_target}D we report experimental and numerical snapshots at $t=\SI{55}{ms}$ when the structure is excited with the signal shown in the inset (with peak displacement $\approx\SI{9}{mm}$).
We find very good agreement between experiments and simulations, with velocity magnitude that is largest in the target region.
Furthermore, to assess the robustness of the focusing performance experimentally, we excite the optimized structure with input signals of varying amplitudes (Fig.~\ref{fig:energy_focusing_single_input_single_target}E).
In Fig.~\ref{fig:energy_focusing_single_input_single_target}F, we present the peak kinetic energy, $\max_t \mT_i$, along the white line in Fig.~\ref{fig:energy_focusing_single_input_single_target}D for the three considered inputs.
Notably, there is a clear concentration of kinetic energy at the specified target region (green shaded area), affirming the efficacy of the focusing mechanism.

This example of energy focusing design demonstrates how our strategy can discover geometries that channel energy towards a desired location without requirements on energy distribution anywhere else.
More control over the spatial localization of energy can be achieved by generalizing this problem to multiple focusing locations.
With such strategy, we can design metamaterial architectures capable of splitting the input energy among specified target regions of space and desired splitting ratio (\href{SI-link}{Supporting Information}, Section~3C).
Additional results on the `dual' problem of energy focusing at a single target area when the metamaterial is excited at different independent locations are reported in \href{SI-link}{Supporting Information}, Section~3D (Fig.~S5, Movie~S5).
Furthermore, results demonstrating that our framework is applicable to arbitrary geometric spaces are reported in \href{SI-link}{Supporting Information}, Section~3F, showing the energy focusing design of a non-periodic metamaterial with a Kagome-like topology (Fig.~S7, Movie~S6).

\subsection*{Reprogramming focusing location}
In Fig.~\ref{fig:energy_focusing_single_input_single_target}, we identified a design capable of concentrating the input energy at a specified location.
Real-world challenges, however, often necessitate tunability of the task.
Crucially, we can harness the high deformability of metamaterials to achieve such reprogrammability within a single architecture.
%
\begin{figure}[htb]
    \centering
    \includegraphics[width=\linewidth]{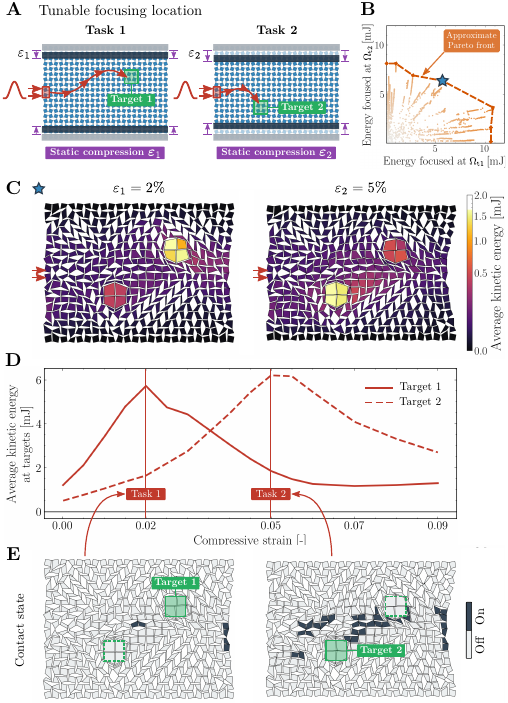}
    \caption{
        \textbf{Reprogramming focusing location.}
        (\textbf{A}) On-the-fly tunability of the focusing location is achieved with a single architecture by harnessing static pre-compression.
        (\textbf{B}) $J^{(\varepsilon_1)}_{\Omega_{\text{t}1}}/t_\text{f}$ and $J^{(\varepsilon_2)}_{\Omega_{\text{t}2}}/t_\text{f}$ for all designs visited during optimization (gray points).
        The resulting Pareto front is highlighted by the orange dashed line.
        (\textbf{C}) Spatial distribution of time-averaged kinetic energy at the two desired compression levels $\varepsilon_1=2\%$ and $\varepsilon_2=5\%$ for the best equi-performance design (highlighted by the blue star marker in panel~B).
        (\textbf{D}) Evolution of the average kinetic energy at $\Omega_{t1}$ and $\Omega_{t2}$ as a function of the applied pre-compression.
        (\textbf{E}) Spatial distribution of contacts at the two desired compression levels $\varepsilon_1=2\%$ and $\varepsilon_2=5\%$.
    }
    \label{fig:switching_focusing_targets}
\end{figure}

To illustrate this concept, we search for a metamaterial that can selectively focus energy in two distinct locations depending on the level of applied pre-compression.
Specifically, we seek a ${24\times18}$-unit design that can focus energy in the target region~$\Omega_{\text{t}1}$ when vertically pre-compressed by $\varepsilon_1=2\%$ and in the region~$\Omega_{\text{t}2}$  when the pre-compression is increased to $\varepsilon_2=5\%$ (Fig.~\ref{fig:switching_focusing_targets}A).
To identify such architecture, we maximize the following objective function
\begin{equation}
    \label{eq:multi_objective_focusing}
    J = w_1 J_1 + w_2 J_2
\end{equation}
where $J_i=J^{(\varepsilon_i)}_{\Omega_{\text{t}i}}$ ($i=1,\,2$) is given by~\eqref{eq:objective_focusing} computed on the structure pre-compressed by $\varepsilon_i$, and ${w_i\geq 0}$ denotes the weight associated with the $i$-th objective.
Each objective $J^{(\varepsilon_i)}_{\Omega_{\text{t}i}}$ is computed through a simulation involving two steps.
In the initial step, a quasi-static compressive strain $\varepsilon_i$ is applied by gradually displacing the top and bottom rows of the structure vertically.
This is followed by a subsequent step where a dynamic pulse, defined by \eqref{eq:pulse_excitation} with ${A=\SI{7.5}{mm}}$ and ${f=\SI{30}{Hz}}$, is applied to the left edge (as indicated by the red arrows in Fig.~\ref{fig:switching_focusing_targets}A).
To systematically sample the Pareto front of this multi-objective problem, we maximize~\eqref{eq:multi_objective_focusing} with $w_1\in[0,1]$ and $w_2 = 1-w_1$ and track $J_i$ for each design in the optimization.
Note that, to enhance the resolution of the Pareto front, we consider 5 initial designs with $\theta_0\in[15, 35]\si{\degree}$ for each weight combination (\href{SI-link}{Supporting Information}, Section~2B for more details on the Pareto sampling).

In Fig.~\ref{fig:switching_focusing_targets}B, we report $J^{(\varepsilon_1)}_{\Omega_{\text{t}1}}/t_\text{f}$ and $J^{(\varepsilon_2)}_{\Omega_{\text{t}2}}/t_\text{f}$ for all the designs visited during optimization (gray points).
The resulting Pareto front (orange dashed line) delineates a set of optimal physically realizable designs that exhibit diverse performance trade-offs in focusing at the two target locations.
Additionally, it demonstrates that our optimization can always improve the performance of the starting design.
In Fig.~\ref{fig:switching_focusing_targets}C, we focus on a design on the sampled Pareto front for which $J^{(\varepsilon_1)}_{\Omega_{\text{t}1}}\approx J^{(\varepsilon_2)}_{\Omega_{\text{t}2}}$ (highlighted by the blue star marker in Fig.~\ref{fig:switching_focusing_targets}B) and plot the time-averaged kinetic energy for $\varepsilon_1=2\%$ (left) and $\varepsilon_2=5\%$ (right).
The two energy maps clearly show the switch between the two tasks at the desired pre-compression levels.
Additionally, in Fig.~\ref{fig:switching_focusing_targets}D we report the time-averaged kinetic energy at $\Omega_{t1}$ (continuous red line) and $\Omega_{t2}$ (dashed red line) as a function of the applied pre-compression for $\varepsilon\in[0, 9]\%$.
We find that as the applied pre-compressive strain is increased, the focused kinetic energy smoothly shifts from $\Omega_\text{t1}$ (continuous red line) to $\Omega_\text{t2}$ (dashed red line).
Interestingly, we find that the optimization algorithm exploits contact between the quadrilateral units to achieve such shift.
As shown in Fig.~\ref{fig:switching_focusing_targets}E, at $\varepsilon=2\%$ only two pairs of units are in contact.
In contrast, at $\varepsilon=5\%$, not only more units get in contact, but they also form a contact chain that connects $\Omega_\text{t1}$ and $\Omega_\text{t2}$.
The added stiffness of this contact chain limits the motion of $\Omega_\text{t1}$ and ultimately causes the incoming energy to deflect towards the lower part of the domain effectively exciting $\Omega_{\text{t}2}$ (Movie~S2).
Remarkably, the full differentiability of our model, which accounts for contact, allows us to discover such contact-based switching mechanism in an automated fashion.


\subsection*{Reprogramming functionality}
\begin{figure*}[htb]
    \centering
    \includegraphics[width=\textwidth]{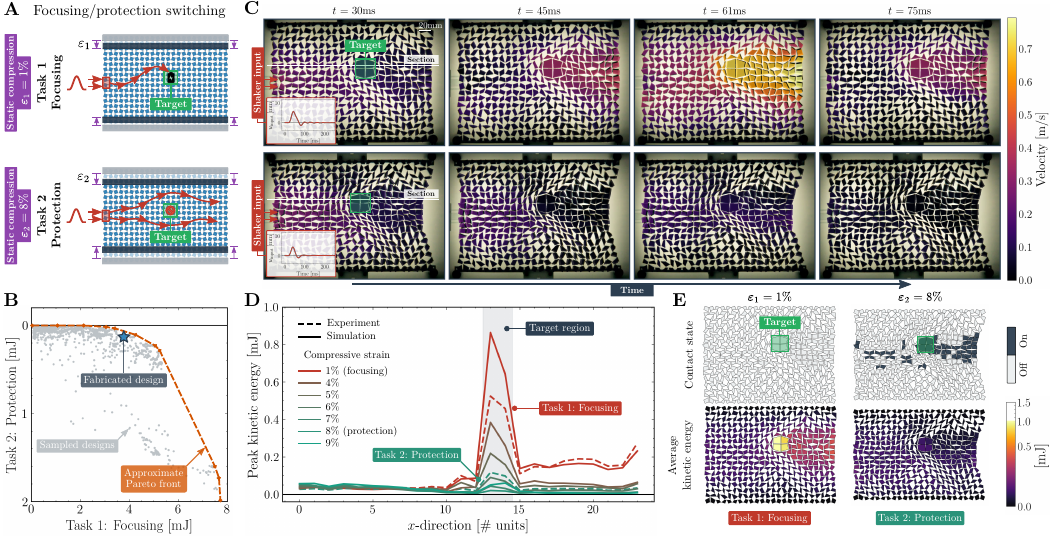}
    \caption{
        \textbf{Reprogramming functionality.}
        (\textbf{A}) We search for a metamaterial architecture with the ability to maximize kinetic energy at a target location for an applied pre-compression of $\varepsilon_1=1\%$ and minimize it for $\varepsilon_2=8\%$.
        (\textbf{B}) All designs visited during optimization (gray points) plotted in the objective space $J^{(\varepsilon_1)}_{\Omega_\text{t}}/t_\text{f}$--$J^{(\varepsilon_2)}_{\Omega_\text{t}}/t_\text{f}$.
        The resulting Pareto front is highlighted by the orange dashed line.
        (\textbf{C}) Experimental snapshots of the fabricated design (marked by a blue star in panel~B) when subject to a pulse-like excitation applied by a low-frequency shaker at $\varepsilon_1=1\%$ (top) and $\varepsilon_2=8\%$ (bottom).
        (\textbf{D}) Peak kinetic energy profiles on a horizontal section across the target region (while line in panel~C) for different levels of applied pre-compression.
        (\textbf{E}) Simulated spatial distribution of contacts (top row) and corresponding kinetic energy distribution (bottom row) at the two desired compression levels $\varepsilon_1=1\%$ and $\varepsilon_2=8\%$.
    }
    \label{fig:switching_focusing_protection}
\end{figure*}

The results of Fig.~\ref{fig:switching_focusing_targets} demonstrate that our design strategy can discover metamaterials with programmable focusing location.
Here we pose a more challenging question: Is our design space rich enough to allow a single metamaterial structure to perform antagonistic tasks?
To answer this question, we deploy our design strategy to search for a metamaterial architecture with the ability to maximize kinetic energy at a target location~$\Omega_\text{t}$ for an applied pre-compression of~${\varepsilon=\varepsilon_1}$ and minimize it for~${\varepsilon=\varepsilon_2}$  (Fig.~\ref{fig:switching_focusing_protection}A).
This multi-task problem is solved maximizing the objective function of~\eqref{eq:multi_objective_focusing} with~${\Omega_\text{t1}=\Omega_\text{t2}=\Omega_\text{t}}$, ${w_1\in[0,1]}$ and ${w_2=w_1-1\in[-1,0]}$ for the tasks of energy maximization and minimization, respectively.
Fig.~\ref{fig:switching_focusing_protection} displays results for a domain consisting of $24\times18$ units under the identical excitation as considered in Fig.~\ref{fig:switching_focusing_targets}, with~${\varepsilon_1=1\%}$ and~${\varepsilon_2=8\%}$.
The Pareto front sampling for this problem is shown in Fig.~\ref{fig:switching_focusing_protection}B.
It is noteworthy that the majority of the explored designs (gray markers) demonstrate a combination of low-focusing and high-protection capabilities.
This suggests that, within the considered metamaterial design space and for the considered target location, it is comparatively easier to design for energy minimization than for energy maximization.

To verify the robustness of the identified optimal designs, we fabricate and test a design on the Pareto front with focusing-to-protection performance ratio of about 25 (i.e.,~${J^{(\varepsilon_1)}_{\Omega_\text{t}}\approx 25 J^{(\varepsilon_2)}_{\Omega_\text{t}}}$, see blue star marker in Fig.~\ref{fig:switching_focusing_protection}B).
We use linear stages connected to the units at the top and bottom rows to apply the desired level of pre-compression and then dynamically excite the pre-loaded sample using a low-frequency shaker.
In Fig.~\ref{fig:switching_focusing_protection}C we report experimental snapshots of the sample pre-compressed by~${\varepsilon_1=1\%}$ (top) and~${\varepsilon_2=8\%}$ (bottom) after dynamic excitation with each quad colored according to the measured velocity magnitude.
In full agreement with the predictions of the model, we find a high velocity region around the target when the sample is pre-compressed by $\varepsilon_1$ with a peak of \SI{0.72}{m/s} observed at the target at $t=\SI{61}{ms}$ (see also Movie~S3 reporting the experimental and simulated dynamic response of this optimized structure).
By contrast, the velocity remains lower than \SI{0.41}{m/s} within the entire domain when the structure is pre-compressed by $\varepsilon_2$.
The efficiency of the design in transitioning from focusing to protection as the applied pre-compression increases is further apparent in Fig.~\ref{fig:switching_focusing_protection}D.
Here, we plot the peak kinetic energy extracted along a horizontal section across the target region for varying levels of applied pre-compression.
In addition to a good agreement between experiments (dashed lines) and simulations (solid lines), the data indicate a consistent decrease in peak kinetic energy within the target region (shaded gray area) as the applied pre-compression increases.
Notably, for $\varepsilon>8\%$, the peak kinetic energy is markedly low.
In this instance as well, we observe that the optimization algorithm utilizes contact to switch between the two tasks (Fig.~\ref{fig:switching_focusing_protection}E).
At $\varepsilon=\varepsilon_1$, no units are in contact.
However, at $\varepsilon=\varepsilon_2$, the units in the target region and numerous units behind it come into contact, resulting in a `jammed' state.
We can interpret the formation of these jammed regions as the attempt of the optimization to locally tune the effective mass density and stiffness to effectively engineer a switch in the dynamic response.
In fact, the jammed areas correspond to the regions that, compared to the rest of the domain, undergo a sharp transition from a high kinetic energy at $\varepsilon_1$ to a low kinetic energy response at $\varepsilon_2$ (Fig.~\ref{fig:switching_focusing_protection}E bottom row).


\section*{Discussion and outlook}
\begin{figure}[htb]
    \centering
    \includegraphics[width=\linewidth]{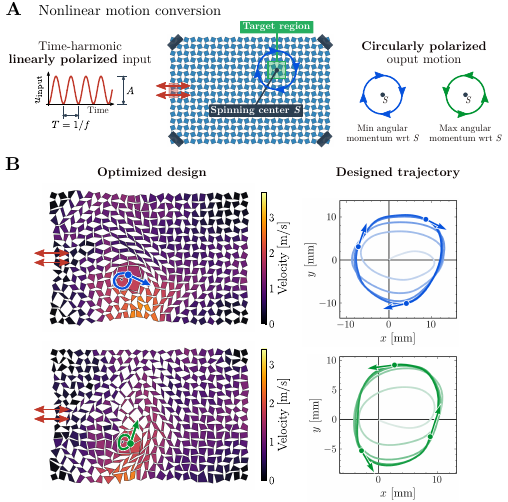}
    \caption{
        \textbf{Nonlinear motion conversion.}
        (\textbf{A}) Circularly polarized motion can be encoded in a metamaterial architecture via our design strategy by optimizing the angular momentum of a target region with respect to a specified point $S$ (midpoint of the target region).
        (\textbf{B}) Two optimized designs (colored according to the velocity map) for clockwise (blue) and counterclockwise (green) output motion.
    }
    \label{fig:spinning_design}
\end{figure}


This study has introduced an automated design framework for identifying non-periodic metamaterials capable of intricate nonlinear dynamic tasks.
Crucially, the resulting optimized architectures feature non-trivial geometries that could not be guessed or intuited \emph{a priori}, highlighting the potential of our design framework for discovering unprecedented material responses in an automated fashion.
Moreover, we have demonstrated the extensibility of this framework to include reprogrammability, facilitating encoding and switching between multiple tasks by leveraging the inherent flexibility of the structures.

While our emphasis in this work has centered on converting large amplitude pulses into controlled energy flow within metamaterial domains, it is worth noting that the framework can be readily extended to accommodate various types of excitation signals and to identify architectures that support a broad spectrum of functionalities.
As an example, in Fig.~\ref{fig:spinning_design} we consider the design of a metamaterial that transforms a linearly polarized large-amplitude harmonic input into an approximately circularly polarized motion at a specified target region.
An architecture capable of counterclockwise motion is identified by maximizing the angular momentum of a target region with respect to a desired point.
In contrast, achieving clockwise motion involves minimizing the angular momentum (see \href{SI-link}{Supporting Information}, Section~3E for details, and Movie~S4).

All together, the presented results indicate that the proposed framework holds promise in identifying material structures capable of complex transient as well as steady-state dynamic behaviors in response to simple actuation inputs.
Moreover, the reprogrammability of such behaviors could be further augmented by enabling simple task selection strategies through pre-deformation, changes in excitation frequency, or variations in loading location.
In conclusion, we envisage that the resulting design paradigm can turn mechanical metamaterials into a rich robotic matter platform for generating soft material embodiments with reconfigurable functionalities.






\subsection*{Data Availability}
All the source code and data are available on GitHub at \href{https://github.com/bertoldi-collab/DifFlexMM}{github.com/bertoldi-collab/DifFlexMM} or provided in the \href{SI-link}{Supporting Information}.

\acknow{
We thank Adel Djellouli for helping set up the dynamic experiments.
We thank Daniel W\"alchli for fruitful discussions on nonlinear optimization and Pareto front search.
We also thank Ben Spoettling for initial efforts on the image tracking code.
S.J. acknowledges funding from Otto Moensted Foundation and William Demant Foundation.
C.B. acknowledges funding from an ETH Zurich Doc.Mobility Fellowship.
The authors gratefully acknowledge support via NSF award numbers 2041440, 2118201, and 2007278.
}
\showacknow{} 

\bibliography{energy_manipulation}

\end{document}